\documentstyle[12pt,epsf]{article}
\def\bi{\bigskip}
\def\noi{\noindent}
\def\be{\begin{equation}}
\def\en{\end{equation}}
\def\bq{\begin{eqnarray}}
\def\eq{\end{eqnarray}}
\def\bc{\begin{center}}
\def\ec{\end{center}}
\def\beit{\begin{itemize}}
\def\eit{\end{itemize}}

\begin{document}
\bc

{\large \bf $a_0(980) \to\gamma\gamma$ and $f_0(980) \to\gamma\gamma$: a 
consistent description \footnote{Work supported by CONACyT 
under contracts 3979P-E9608, J2764-E}}

\ec
\vspace{1.5cm}
\bc
{ \bf J.L. Lucio M. \footnote{email: lucio@ifug1.ugto.mx} and 
M. Napsuciale \footnote{email: mauro@ifug4.ugto.mx} }\\[2.3cm]
{\it Instituto de F\'\i sica, Universidad de Guanajuato}\\
{\it Lomas del Bosque \# 103, Lomas del Campestre}\\
{\it 37150 Le\'on, Guanajuato; M\'exico}\\
\ec


\vspace{1.5cm}
\vspace{1.5cm}
\begin{center}
{\bf Abstract.} 
\end{center}

\bi
\bi
\bi

We work out the  Linear Sigma Model (LSM) predictions for the $2 \gamma$ decay 
rates of the $a_0(980),~f_0(980)$ mesons under the assumption that 
they are respectively the $I=1$ and $I=0$  members of the $\bar q q$ scalar 
nonet. Agreement with experimental data is achieved provided we include 
the contribution of a $\kappa$ meson with  mass 
 $\approx 900 MeV$, and  a scalar mixing angle 
($\sigma -f_0$ mixing in the $\{|NS>,|S>\}$ basis) 
$\varphi_s\approx -14^\circ$, as predicted by the model.

\bi

\setlength{\baselineskip}{1\baselineskip}

\newpage

\noi {\bf  Introduction.} 

\bi 

The  scalar is the most controversial sector of low energy QCD. 
In contrast with the pseudoscalars or vector mesons where the corresponding 
multiplets have been unambiguously established and the hadron properties can be
interpreted in terms of constituent quarks or effective theories for low 
energy QCD, the scalar meson identification faces severe problems. 

The Particle Data Group (PDG) [1] candidates for the 
ground state $\bar q q$ scalar  nonet are : the $f_0(980)$, $f_0(1370)$ and 
the recently resurrected  $f_0(400-1250)~(\sigma ?)$ meson  for two sites in 
the  $I=0$ sector; the $a_0(980)$ and $a_0(1450)$ for the isovector 
scalar meson, and the 
$K^*_0(1430)$ for the isospinor scalar meson.

Over the past years 
experimental evidence has accumulated for the existence of  light scalar 
mesons [2-4]. A reanalysis of data [3] which introduces a  phenomenological 
background phase shift ($\delta_B$) claims the existence of an isovector 
$\kappa (\approx 900)$ and a light $\sigma (\approx 600)$ isoscalar meson. 
This phase shift can be naturally interpreted in terms of four-meson 
interactions within the Linear Sigma Model (LSM). Alternative analysis of the 
same data [4] arrived to the same conclusion.  
There exist claims  for the existence of an even  lighter isoscalar meson 
$\sigma ( 400-600)$ in different contexts [4-6], and  the existence of two 
scalar meson nonets has also been suggested [7].

The most important drawback for 
the identification of the $a_0(980)$ and $f_0(980)$ as the $\bar q q$ scalar 
isovector and isosinglet respectively, is 
their tiny coupling to two photons. 
The PDG  quotes the averaged values 
$\bar\Gamma (a_0(980)\to 2\gamma)\times BR(a_0(980)\to \pi^0\eta)= 
0.24^{+0.08}_{-0.07}~ KeV $ and 
$\bar\Gamma (f_0(980)\to 2\gamma)= 0.56\pm 0.11 KeV$ , as reported by the JADE 
[8], and Crystal Ball [9] collaborations. In the case of the 
$f_0(980)$, the experimental result is averaged by the PDG with an estimate 
by Morgan and Pennington [10]. 

\bi

On the theoretical side, a large amount of work has been done trying to 
understand the structure of the $a_0(980)$ and $f_0(980)$ mesons. There 
exist calculations for the $a_0(980),f_0(980) \to 2\gamma$ decays using a  
variety of approaches [11-13], in particular, in  different versions 
of the quark model [11]. The generally accepted conclusion, 
is that the $a_0(980), f_0(980) \to\gamma\gamma$ decay widths are not 
consistent with a $q\bar q$ structure.  Thus, other possibilities 
such as a molecule picture [12] 
and a $\bar q q \bar q q $ structure [13] have been explored and found to 
be consistent with the tiny coupling to two photons.   

Recent data from Novosibirsk [14] and forthcoming experiments at high 
luminosity $\phi$ factories, will shed some light on this controversial 
sector. Eventually, the precise measurement of the two photon decay of the 
 $a_0(980)$ and  $f_0(980)$ could discriminate among the various proposals 
for the lowest lying $\bar q q$ scalar nonet.

Recently, scalar meson properties were studied in a Linear Sigma Model 
which incorporates a t'Hooft interaction [6] . The model predicts that the 
members 
of the  scalar nonet are: $\{\sigma(\approx 400), f_0(980),
\kappa (\approx 900)$ and $a_0(980)\}$, with a scalar mixing angle (in the 
$\{|NS>,|S>\}$ basis) $\phi_s \approx -14^\circ$. 
In this work we pursue the study of the implications of the LSM for the 
scalar mesons phenomenology by computing the $ a_0 (980),f_0(980) \to 
\gamma\gamma $ transitions within the model.

\bi
\bi
\bi

\noi {\bf Meson loop contributions to $S \to\gamma\gamma$}

\bi

Lorentz covariance and gauge invariance dictates the most general form of 
the $S\to\gamma\gamma$ transition amplitude ($S$ denoting a  scalar meson) : 
\be
{\cal M}(S\to \gamma (k,\epsilon)~\gamma(q ,\eta))= 
{i\alpha\over \pi f_{_K}}V^{^S}~ (g^{\mu\nu} q 
\cdot k-k^\mu q^\nu)\eta_\mu \epsilon_\nu  .
\en

The charged meson (hereafter denoted $M$) loop contributions to 
$ S \to\gamma\gamma$  are depicted in Fig.(1).
A straightforward calculation yields

\be
V^{^S}_{_M}=\frac{2~f_{_K}~g_{_{SMM}}}{m^2_{_S}}\left[-\frac{1}{2}
+\xi^{^S}_{_M} I (\xi^{^S}_{_M})\right] ,
\en
\noi where  $\xi^{^S}_{_M}= {m^2_{_M}\over m^2_{_S}}$, and $I(x)$ denotes 
the loop integral

\bq 
I (x)=\left\{ \begin{array}{lll} 2 \Big ({\rm Arc \, sin} 
\sqrt{\frac{1}{4x}}\Big)^2 & x > \frac{1}{4}\\
2\Big({\frac{\pi}{2}}+i\ln\Big(\frac{1}{\sqrt{4x}}+ 
\sqrt{\frac{1}{4x}-1} \Big)\Big)^2  & x < \frac{1}{4} . \qquad
\end{array}\right. 
\eq

\bi
\noi The decay width is given by:
\be
\Gamma (S \to \gamma\gamma) =\frac{\alpha^2}{64\pi^3} 
                    \frac{m^3_{_S}}{f^2_{_K}}|V^{^S}|^2.
\en

\bi

\vskip2ex
\centerline{
\epsfxsize=250 pt
\epsfbox{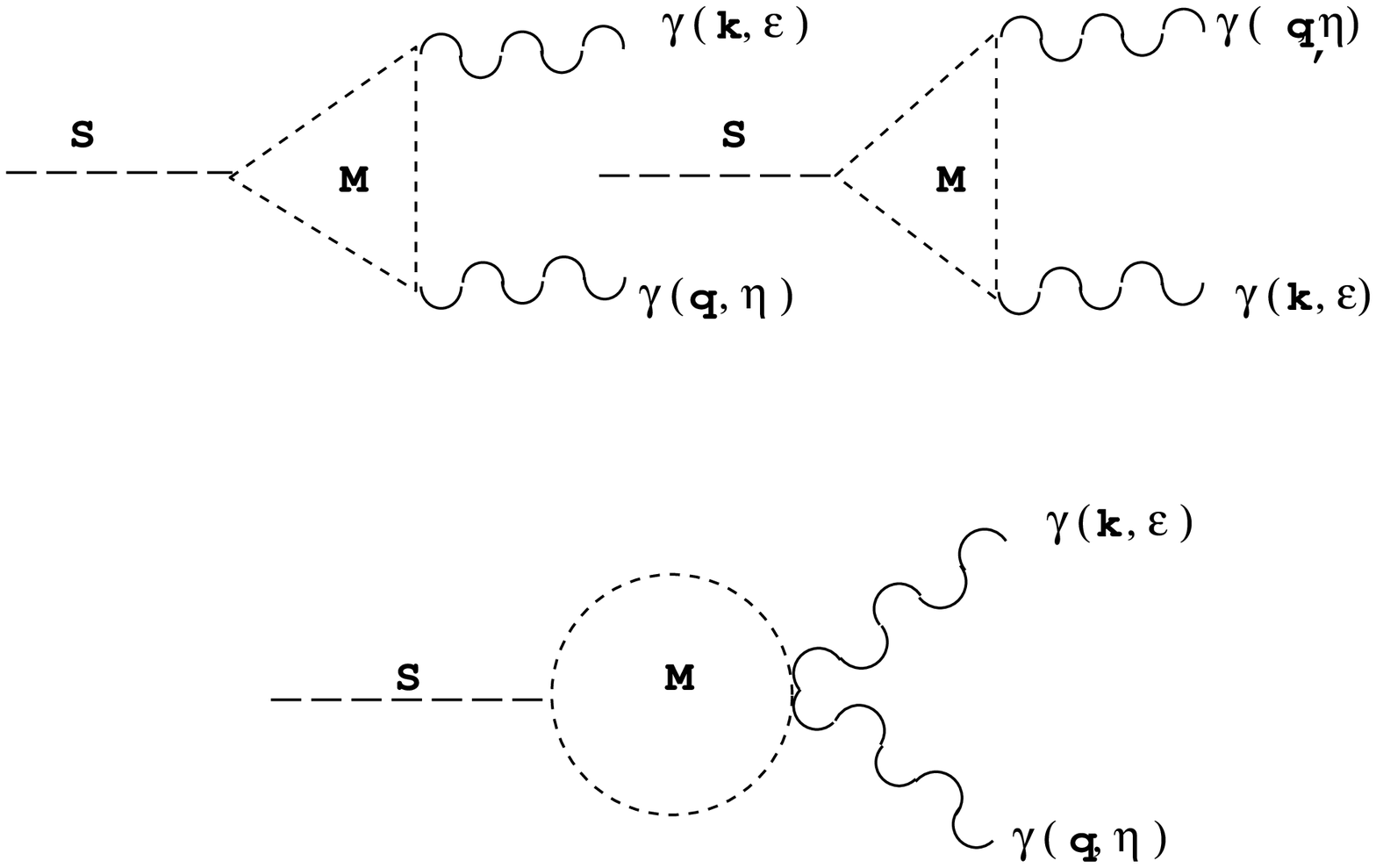}}
\vskip4ex
\bc
{\small{Fig.1}\\
 Charged meson loop contribution to $S \to \gamma\gamma$.}
\ec

\bi
\noi Let us first analyze the $a_0\to\gamma\gamma$ transition. The main 
contribution to this decay comes from a loop of charged $K$'s. The $a_0KK$  
coupling constant is dictated by chiral symmetry [5,6]

\be
g_{_{a_0 K^+ K^-}} = \frac{m^2_{a_0}-m^2_{_K}}{2f_{_K}}.
\en

Using  the values reported by the PDG for the $a_0,~K$ masses and $f_{_K}$, 
we obtain:
\be
 V^{a_0}_{_K} = 0.42. 
\en
This is to be compared with the experimental result [1]

\be
|V^{a_0}_{exp}|= 0.34 \pm 0.05, 
\en

\noi extracted from the PDG average  by assuming $BR(a_0\to\pi^0\eta)=1$.
Within the model we are considering, the only other contribution arises from 
an isovector scalar meson loop, with the corresponding coupling constant 
dictated also by chiral symmetry [6]:

\be
g_{a_0 \kappa\kappa} = -\frac{m^2_{a_0} -m^2_\kappa}{2 f_\kappa}= 
-{m_{a_0}^2-m^2_\kappa \over 2(m^2_{_K}f_{_K}-m^2_\pi f_\pi)}m^2_\kappa .
\en

It is worth noticing the minus sign in Eq.(8). The crucial sign 
difference between $SPP$ and $SSS$ is generated  via the chiral structure 
${\cal S}+i{\cal P}$  entering in the construction of the chirally 
symmetric Lagrangian [6] (${\cal S}$ and ${\cal P}$ denoting the 
pseudoscalar and scalar nonets respectively). Clearly, $K$ and $\kappa$ meson 
contributions will interfere destructively whenever the $\kappa$ meson mass 
lie below the $a_0(980)$ meson mass. 

The model predicts $m_\kappa \approx 900 MeV$ [6]. Using this value we obtain:
\be
V^{a_0}_{_{LSM}}=0.36
\en
in good agreement with experimental results in Eq.(7).

\bi

We can reverse the argument. Considering one standard deviation in the 
experimental data, the $\kappa$ meson mass is constrained by the 
$a_0(980)\to 2\gamma$ decay to lie in the range $m_\kappa \in [820,935]~MeV$. 
The central value in Eq.(7) correspond to  $m_\kappa =870 MeV$.
In this respect, it should be mentioned that a kappa mass of 887 MeV was found 
by Svec and collaborators in ref. [2]. More recently, independent 
reanalysis of $K\pi$ phase shifts [3,4], conclude 
$m_\kappa \approx 900 MeV$, whereas theoretical analysis lead to the same 
conclusion [4-7].
   
\bi

The $f_0 (980)\to\gamma\gamma$ decay can be treated in analogy to  
$a_0 (980) \to\gamma\gamma$. In this case, however, the mixing between the 
$\sigma ~(f_0 (400-1200)?)$ and $f_0(980)$ must be taken into account. The 
invariant amplitude describing the process is given by Eqs.(1,2) with $S=f_0$.
 The observed decay rate

\be
\Gamma(f_0 \to\gamma\gamma) = \frac{\alpha^2}{64\pi^3} \frac{m^3_{f_0}}
{f^2_{_K}}|V^{f_0}_{exp}|^2 \cong 0.56 \pm 11 KeV ,
\en  

\noi requires 

\be
|V^{f_0}_{exp}| = 0.53\pm 0.05.
\en

In the model we are considering, the $f_0(980) \to\gamma\gamma$ transition 
gets contributions from  loops of scalar and pseudoscalar mesons 
$(K, \kappa ,\pi)$. Calculations for the 
amplitude in this case yields 

\be
V^{f_0}_{_M} = \Bigg(\frac{2f_{_K}g_{_{f_0MM}}}{m^2_{f_0}} \Bigg) 
(-{1\over2}+ \xi^{f_0}_{_M}~I (\xi^{f_0}_{_M} )) ,
\en

\bi

\noi where $\xi^{f_0}_{_M} = \Big( \frac{m_{_M}}{m_{f_0}} \Big)^2$ , 
$M=\pi,K,\kappa$ and $I(\xi^{f_0}_{_M})$ is given by Eq. (3). 
Using the PDG values [1] for the $K$ and $\pi$ masses and 
$m_\kappa = 870 MeV$, as required by the the central value of the 
$a_0(980) \to 2 \gamma$ decay in Eq.(7), we obtain 

\be
V^{f_0}_{_M} =f_{_K} \Bigg(\frac{g_{_{f_{0}MM}}}{m^2_{f_0}} \Bigg)N_{_M}
\en
with  $N_{_K} = 1.06$, $N_\kappa =0.12$, $N_\pi =(- 1.10+0.48 i)$. 

\bi

We still must fix the $f_0$ couplings which are affected by its mixing with 
the $\sigma$ meson. The physical $\sigma, f_0$ fields are related to the 
$\{|S>,|NS>\}$ isoscalar fields by [5,6]

\bq
|\sigma >=cos(\phi_s) |NS>~ -~ sin(\phi_s) |S>, \\ \nonumber
|f_0>=sin(\phi_s) |NS>~ +~ cos(\phi_s) |S>,
\eq

\noi in such a way that in the zero mixing limit the $f_0(980)$ is purely 
strange. The scalar mixing angle has been estimated to be 
$\phi_s\approx -14^\circ$ [6]. Thus, the physical $f_0(980)$ is mostly 
strange. The more conventional mixing angle $\theta_s$ in the octet-singlet   
basis is related to $\phi_s$ through  $\theta_s = \phi_s - arctan(\sqrt{2})$. 

In the zero mixing limit, the model predicts
\be
g_{_{f_0MM}}(\phi =0)= {m^2_{f_0}-m^2_{_M} \over 2f_{_M}}.
\en 

To leading order in the mixing angle we can use
\be
g_{_{f_0MM}}= g_{_{f_0MM}}(\phi =0)~ F_{_M},
\en 

\noi where $  F_{_M} $ stand for the mixing factors
\be
F_\pi = sin(\phi_s),~~~ F_{_K}=F_\kappa= sin(\phi_s) + \sqrt{2} cos(\phi_s).
\en
Pion and kappa loop contributions are suppressed by mixing factors and the 
large kappa mass respectively. Thus, again, kaon loops dominates the 
$f_0(980)\to \gamma\gamma$ transition. Numerically, using 
$\phi_s=-14^\circ$, Eq.(13) yields $V^{f_0}_{_K}=0.44$, 
whereas taking all contributions into account we obtain

\be
|V^{f_0}_{_{LSM}}|=0.52,
\en

\noi to be compared  with experimental data in Eq.(11). Thus, although $\pi$ 
and $\kappa$ 
contributions are small, they are necessary in order to achieve consistency 
with the experimental results. Again, the argument 
can be turned around. The scalar mixing angle is constrained by the 
experimental errors to lie in the range [$-30^\circ,-5^\circ$], the central 
value in Eq.(11) corresponding to $\phi_s = -16^\circ$

\bi

{\bf Summary.}

\bi

Summarizing, we compute the $a_0(980), f_0(980) \to 2\gamma$ 
decay rates in the framework of a LSM, assuming that the $a_0(980), f_0(980)$ 
mesons are the isovector and isoscalar members of the $\bar q q$ nonet.
The two photon decays are induced by loops of charged mesons, the 
dominant contribution arising from a loop of charged $K$ mesons. Agreement 
with the experimental data is achieved for a $\sigma -f_0$ mixing (in 
the $\{|NS>,|S>\}$ basis) $\phi_s= -16^\circ$, and a $\kappa$ mass 
$m_\kappa =870~MeV$. The required mixing angle and $m_\kappa$ are 
consistent with the values predicted by the model [6].

\bi

{\bf Acknowledgments.}

\bi
We wish to thank  M. D. Scadron for an initial collaboration on this topic and 
useful discussions.

\vfill
\newpage

{\bf References.}

\begin{itemize}
\item [1.-] Particle Data Group, Eur. Phys. Jour. {\bf C3} (1998).

\item [2.-] D. Alde et.al. GAMS Coll. Phys.Lett. B397, 350 (1997); 
M. Svec, Phys. Rev. D 45, 1518 (1992).
 
\item [3.-] S.Ishida, et.al. Prog. Theor. Phys. 96, 745 (1996); 
S. Ishida, hep-ph/9712229; T. Ishida et.al. hep-ph/9712230; 
M. Ishida and S. Ishida hep-ph/9712231; K. Takamatsu et.al. hep-ph/9712232: 
M. Ishida, S. Ishida and T. Ishida hep-ph/9712233.

\item [4.-] F. Sannino and J. Schechter Phys.Rev D52,96 (1995) ; M. Harada, 
F. Sannino and J.Schechter Phys Rev D54, 1991 (1996); D. Black et.al. Phys 
Rev.D58, 054012, (1998);J. A. Oller and E. Oset  
Nucl. Phys. A620, 438,(1997) hep-ph/9702314; L. Lesniak, hep-ph/9807539.

\item [5.-] M. D Scadron, Phys Rev. D26, 239 (1982).

\item [6.-] M. Napsuciale hep-ph/9803396. M. Ishida hep-ph/9902260.

\item [7.-]  E. Van Beveren et.al Z.Phys. C30, 615 (1986); E. Van Beveren 
and G. Rupp  hep-ph/9806246, hep-ph/9806248. 

\item [8.-] T. Oest et.al, JADE Coll. Z. Phys.C47, 343 (1990).

\item [9.-] D. Antreasyan et.al. Crystal Ball Coll. Phys. Rev. D33, 1847 
(1986); H. Marsiske et. al.  Crystal Ball Coll. Phys. Rev. D41,  3324 (1990).

\item [10.-] D. Morgan and M. Pennington  Z. Phys. C48, 623  (1990).

\item [11.-] A. Bramon and M.Greco, Lett. Nuovo Cimento 2, 522 (1971);
S.B. Berger and B.T. Feld  Phys. Rev. D8, 3875 (1973): S. Eliezer, J. Phys. 
G1 , 701 (1975); 
J. Babcoc and J. L. Rosner  Phys. Rev. D14, 1286 (1976); M. Budnev and A.E. 
Kaloshin Phys. Lett. 86B, 351 (1979); 
 N.N. Achasov, S.A. Denayin and G. N. Shestakov Z. Phys. C16, 55 (1982);
E. P. Shabalin, JETP Lett. 42, 135 (1985); J. Ellis and J. Lanik, 
Phys. Lett. 175B, 83 (1986); S. Narison, Phys. Lett 175B, 88 (1986); 
C. A. Dominguez and N. Paver Z. Phys. C39, 39 (1988); Z.P. Li, F.E. Close and 
T. Barnes Phys Rev D43, 2161 (1991); 
A. S. Deakin et. al. Mod. Phys. Lett. A9, 2381 (1994);
 
\item [12.-]  J. Wenstein and N. Isgur, Phys Rev. D27, 588 (1983); T. Barnes 
Phys. Lett. 165B, 434 (1985).
 
\item [13.-]  N.N. Achasov, S.A. Denayin and G. N. Shestakov Phys. Lett. 
108B, 134 (1982); Z.Phys. C16, 55 (1982); E. P. Shabalin, Sov. J. Nucl. 
Phys. 46, 485 (1987); N.N. Achasov and G. N. Shestakov Z.Phys. C41, 309 
(1988); N.N. Achasov and V.N. Ivanchenko, Nucl. Phys. B315, 465 (1989): 
N.N. Achasov hep-ph/9803292.

\item [14.-] N. N. Achasov and V. V. Gubin, Phys.Atom.Nucl.61:1367,(1998); 
V.M. Aulchenko et. al. Phys.Lett. 436B, 199 (1998); M.N. Achasov et. al. 
hep-ex/9809013; M.N. Achasov et.al. Phys.Lett.438B, 441 (1998);  M.N. Achasov 
et.al.JETP Lett. 68, 573 (1998);V.M. Aulchenko et. al. 
Phys.Lett. 440B, 442 (1998).

\end{itemize}

\end {document}